\documentclass[twocolumn,showpacs,preprintnumbers,amsmath,amssymb]{revtex4}
\usepackage{graphicx}% Include figure files
\usepackage{dcolumn}% Align table columns on decimal point
\usepackage{bm}% bold math

\begin{document}

\preprint{}

\title{Determination of the upper and lower bound of masslimit of degenerate fermionic dark matter objects}% Force line breaks with \\

\author{G\'abor Kupi}
 \email{gabor.kupi@weizmann.ac.il}
\affiliation{%
 Weizmann Institute of Science, PO Box 26, Rehovot 76100, Israel
}%

\date{\today}
\begin{abstract}
We give a gravitational upper limit for the mass of static degenerate fermionic dark matter objects. The treatment we use includes fully relativistic equations for describing the static solutions of these objects. We study the influence of the annihilation of the particles on this mass limit. We give the change of its value over the age of the Universe with annihilation cross sections relevant for such fermions constituting the dark matter. Our calculations take into account the possibility of Dirac as well Majorana spinors.  
\end{abstract}

\pacs{95.30.Sf, 95.35.+d, 97.10.-q}
\maketitle

\section{Introduction}
Existence of a considerable amount of cold dark matter in the present Universe has been regarded as a robust ingredient in recent astrophysics and cosmology. Several hypothetical particles were suggested in order to solve the problem of the cold dark matter. Some of them are bosons: axions, Q-balls\ldots The others are fermions: neutralinos, axinos, gravitinos, quintessino\ldots (One can read a good summary about the non-barionic dark matter candidates in \cite{Fornengo}.) 

We study some general features of the degenerate fermionic matter here. Degeneracy means that every energy states are occupied, that is such a system is spherical symmetric and does not rotate.
Our aim is to determine the maximal possible mass of {\it static} degenerate fermion balls. 

First we make a naive non-relativistic estimation because the non-relativistic system has been studied in the literature for decades. Several papers were devoted to the non-relativistic degenerate fermionic dark matter with or wihtout barionic component (see for example \cite{ArbolinoRuffini,Viollier1,Bilic,Messer,HertelThirring1,HertelThirring2}). Papers whose topic is some aspect of the relativistic behaviour of these systems are rare (\cite{Kupi}). We devote the second part of this paper to giving  an exact relativistic mass limit for these systems. In the third part of the paper we study the influence of the annihilation of the particles on this mass limit.

In non-relativistic case we use polytropic equation of state, in relativistic case we do not use such approximation but the exact equation which one can get for degenerate fermions.

If we assume that the dark matter consists of neutral, weakly interacting fermions, 
their spatial distribution is driven by gravity and Fermi distribution. 
 $kT \ll m_0 c^2$ is valid for such kind of particles today, that is if the
density is sufficiently high, the degenerate ($T=0$ K) Fermi distribution approximates the real distribution 
well. $k$ is the Boltzmann constant, $T$ is their temperature, $m_0$ is their rest mass and $c$ is the speed of light.
One can write that 
\begin{equation} \label{fermi}
z\pi {p_{F}}^3/3 h^3=\varrho/m_0,
\end{equation}  
where $\varrho$ is the mass density of 
particles, $p_{F}$ is their Fermi momentum and $h$ is the Planck constant. 
If the particles are Majorana spinors, the $z$ factor is 8. If Dirac spinors, then 16.

\section{Non-relativistic case}

 We dispense with barionic matter and deal only with one component dark matter. This assumption is not good for cosmological scales but probably suitable for smaller scales. We derive the equations of the system first.
  
We can obtain $p_F$ from (\ref{fermi}): $p_F=h\left(\frac{3}{z\pi}\frac{\varrho}{m_0}\right)^{1/3}.$ The energy density for non-relativistic case is 
\begin{equation} \label{nonrelepsilon}
\epsilon=\frac{z\pi}{h^3}\int_0^{p_F}\!\!\!\epsilon(p)p^2dp=\frac{z\pi}{2h^3 m_0}\int_0^{p_F}\!\!\!p^4dp,
\end{equation}  
 we can integrate it easily $$\epsilon=\frac{z\pi}{10h^3 m_0}p_F^5
=\frac{3h^2}{10m^{8/3}_0}\left(\frac{3}{z\pi}\right)^{2/3}\!\!\!\!\varrho^{5/3}\!\!\!.$$ $P=\frac{2}{3}\epsilon$ is valid for ideal gases. We know the equation of state now:
 \begin{equation} \label{allegy}
 P={\frac{h^2}{5m^{8/3}_0}}\left({\frac{3}{z\pi}}\right)^{2/3}\varrho^{5/3}\ \ (\equiv K \varrho^{5/3}),
 \end{equation}

Navier--Stokes equations are very simple for spherical symmetric matter distribution in hydrostatic 
equilibrium: 
 \begin{equation}\label{NavierStokes}
 \frac{dP}{dr} = -\frac{dU}{dr}\varrho,
 \end{equation}
where $r$ is the distance from the origin and $U$ is the gravitational potential. 

One needs also the Poisson equation 
$ \nabla^2 U=4 \pi G \varrho$, where $G$ is the gravitational constant.
It can be reformulated for spherical symmetric case: 
 \begin{equation} \label{Poisson}
 \frac{d^2 U}{dr^2}+{\frac{2}{r}}\frac{dU}{dr}=4\pi G \varrho.
 \end{equation}
 
If one sub\-stitutes $P$ in (\ref{NavierStokes}) from (\ref{allegy}) and in\-teg\-rates it, then gets re\-la\-tion bet\-ween
$\varrho$ and $U$: $\varrho=\left[{\frac{2}{5K}}(U_0-U)\right]^{3/2}$, where $U_0$ is the constant of integration.  
With this the Navier--Stokes equation is: $ \frac{d^2 (U_0-U)}{dr^2}+{\frac{2}{r}}\frac{d(U_0-U)}{dr}=-4\pi G \left({\frac{2}{5K}}\right)^{3/2}
(U_0-U)^{3/2}$. If we introduce the following variables $x=\sqrt{a} r$ (where $a= \frac{8z \sqrt{2}}{3} \pi^2 G h^{-3} m^4_0$) and $\psi = x(U_0-U)$, then it becomes
very simple:
 \begin{equation} \label{Emden}
\frac{d^2\psi}{dx^2}=-\frac{\psi^{3/2}}{\sqrt{x}}.
 \end{equation}
It is the well known Emden equation for $n=3$.
When one solves this equation, has only one free parameter,  ${\frac{d\psi}{dx}} |_{x=0}$, because $\psi(0)$ must be 0 in order to avoid singularity at $x=0$.

\section{Mass limit in non-relativistic case}

In the non-relativistic case we can make a raw estimation of upper limit of the mass of such a fermion ball. The mass 
density can be calculated from the previous relations $\varrho = \frac{a}{4\pi G} {\left(\frac{\psi}{x}\right)}^{3/2}$. If we put this into the common expression of total mass, we get the following
$$M=4\pi \int_0^{r_{F}}{\varrho(r^{\prime})r^{\prime2}dr^{\prime}}  \Rightarrow$$ 
$$M=\frac{\sqrt{6}2^{3/4}h^{3/2}}{8\sqrt{z}\pi G^{3/2} m^2_0} \int_0^{x_{F}}{\psi(x^{\prime})^{3/2}\sqrt{x^{\prime}}dx^{\prime}},$$
where $r_{F}$ and $x_{F}$ denote the radius of this object, where the mass density drops to zero. We can introduce
$\kappa(x_{F})$ notation for the integral and $A$ for the constant factor excluded the $m_0$. Then we have $M=\frac{A}{m^2_0}\kappa(x_{F})$. We know the general scaling feature of this equation $M \propto \frac{1}{x_{F}^3}$ (see \cite{ViollierTrautmann} for example), that is $\kappa(x_{F})\equiv \frac{B}{x_{F}^3}$, where the numerical value of $B$ is approximately 132.384. If the mass reaches the Schwarzschild mass limit, we get a black hole with $M_{Sch}=\frac{A}{m^2_0}\frac{B}{x_{F}^3}=\frac{r_F c^2}{2G}=\frac{x_F c^2}{2G \sqrt{a}}$. After some algebra one can write relation between $M_{Sch}$ and $m_0$
 \begin{equation} \label{nonrel}
 M_{Sch}= \frac{\sqrt{6}}{8\sqrt{z}\pi} \left(\frac{ch}{G}\right)^{3/2} B^{1/4} \frac{1}{m^2_0}.
 \end{equation}
If $M_{Sch}$ is measured in solar masses and $m_0$ in GeV, it means $ M_{Sch}\approx 2.11 \frac{1}{m^2_0}$ in Dirac case and
 $ M_{Sch}\approx 2.98 \frac{1}{m^2_0}$ in Majorana case.

\section{Relativistic case}

For spherical symmetric case the metric can be written in the form

$$ds^{2} = e^{\nu}c^{2}dt^{2}-r^{2}(d\theta^{2}+sin^{2}\theta d\varphi^{2}) - e^{\lambda}dr^{2},$$
where $\nu$ and $\lambda$ are the function of $r$ and $t$. We study the static case here, that is $\nu=\nu(r)$, $\lambda=\lambda(r)$ \cite{Landau}. 

One can obtain the Einstein equations from this and the stress-energy tensor of the ideal gas for static case:
$$
8\,{\frac {\pi \,GP}{{c}^{4}}}={e^{-\lambda}} \left( {\frac {\nu^{\prime}}{r}}+\frac{1}{{r}^{2}} \right) -\frac{1}{{r}^{2}}
$$
$$
8\,{\frac {\pi \,GP}{{c}^{4}}}=\frac{1}{2}\,{e^{-\lambda}} \left( \nu^{\prime\prime}+\frac{\nu^{\prime}}{2} +{\frac {\nu^{\prime}-{\lambda^{\prime}}}{r}}-\frac{\nu^{\prime}\lambda^{\prime}}{2} \right)
$$
$$
8\,{\frac {\pi \,G  \epsilon  }{{c}^{4}}}=-{e^{-\lambda}} \left( \frac{1}{{r}^{2}}-\frac{\lambda^{\prime}}{r} \right) +\frac{1}{{r}^{2}},
$$
where $P$ is the pressure and $\epsilon = \varrho\,{c}^{2}+3P/2 $ is the energy density of ideal gas. Prime denotes the 
differentiation along $r$. One can derive an explicite ordinary differential equation for $\nu^{\prime\prime}$ from these equations.
\begin{widetext}
\begin{equation} \label{nu2}
 \nu^{\prime\prime} ={\frac {30\,\pi \,GP{r}^{2}\nu^{\prime}+2\,\pi \,GP{r}^{3} \nu^{\prime2}+36\,\pi \,GPr+4\,{c}^{2}\varrho\,{r}^{3}\pi \,G \nu^{\prime2}-{c}^{4} \nu^{\prime2}r+12\,\pi \,G {c}^{2}\varrho\,{r}^{2}\nu^{\prime} -2\,{c}^{4}\nu^{\prime} +8\,\pi \,G {c}^{2}\varrho\,r}{r \left( 8\,\pi \,GP{r}^{2}+{c}^{4} \right) }}
\end{equation}
\end{widetext}

The mass density is the following:
\begin{widetext}
\begin{eqnarray} \label{Density}
&\varrho&=\frac{z\pi}{h^3}\int_0^{p_F}{\sqrt{m_0^2+p^2/c^2}}p^2dp =\notag\\
& &=\frac{z\pi}{8h^3}\left[\left(m_0^2c^2p_F+2p_F^3\right)\sqrt{m_0^2+p_F^2/c^2}-m_0^4c^3ln\left(\frac{p_F}{cm_0}+\sqrt{1+\left(\frac{p_F}{cm_0}\right)^2}\right)\right].
\end{eqnarray}
\end{widetext}

The pressure can be obtained by means of integrating of the current of linear momentum:
\begin{widetext}
\begin{eqnarray} \label{Pressure}
&P&=\frac{z\pi}{3h^3}\int_0^{p_F}mv^2p^2dp = \frac{z\pi}{3h^3}\int_0^{p_F}\frac{cp^4}{\sqrt{m_0^2c^2+p^2}}dp = \notag\\
& &=\frac{z\pi c}{3h^3}\left[\frac{p_F}{4}\sqrt{m_0^2c^2+p_F^2}\left(p_F^2-\frac{3m_0^2c^2}{2}\right)+\frac{3}{8}m_0^4c^4ln\left(\frac{p_F}{m_0c}+\sqrt{1+\left(\frac{p_F}{m_0c}\right)^2}\right)\right].
\end{eqnarray}
\end{widetext}

We can not derive an equation of state with simple (polytrop) form from (\ref{Density}) and (\ref{Pressure}) as in non-relativistic or highly relativistic limit. The best choice is to use $p_F$ as an independent variable instead of $P$ or $\varrho$.
One needs an equation for $p_F$ now. 
It can be derived from the feature of stress-energy tensor $T^k_{i;k}=0$:
\begin{equation} \label{rho1}
\varrho^{\prime} = -\frac{\varrho{c}^{2}+\frac{5}{2}P}{2\frac{dP}{d\varrho}}\nu^{\prime}.
\end{equation}
If one substitutes $P$ and $\varrho$ in (\ref{rho1}), he obtains for $p_F$: 
\begin{equation} \label{Pf}
p_F^{\prime} = -\frac{3h^3}{2\pi z }\frac{\varrho(p_F){c}^{2}+\frac{5}{2}P(p_F)}{p_F^4}\sqrt{m_0^2+p_F^2/c^2}\nu^{\prime}
\end{equation}

The system (\ref{nu2}), (\ref{Pf}) contains two unknowns and two ordinary differential equations, that is it can be solved.  
((\ref{Density}),  (\ref{Pressure}) are only implicite auxiliary equations in technical sense.)
  The calculations themselves follow closely Oppenheimer and Volkov's calculations \cite{Oppenheimer}.
  
This system has a scaling feature which can be seen without solving the equations.
If $\varrho_1$ and $\nu^{\prime}_1$ are solution of the system with $m_1$, then
the following functions also solve the system with $m_2$:
$$\varrho_2 = \varrho_1\left(\frac{m_2}{m_1}\right)^{4},$$
$$\nu^{\prime}_2 = \nu^{\prime}_1\left(\frac{m_2}{m_1}\right)^{2}$$
with the transformation of $r_1$ 
$$r_2 = r_1\left(\frac{m_1}{m_2}\right)^{2}.$$

A consequence of this feature is the following:
\begin{equation} \label{mscaling}
M_2 = M_1\left(\frac{m_1}{m_2}\right)^{2}.
\end{equation}
We see that the total mass is in inverse proportion to the square of the rest mass of the particles even as in non-relativistic case.

In order to solve these equations we need initial conditions. $\nu$ does not occur explicitely in the equations, only its derivatives. (Non-static equations contain $\nu$ itself.) It means that its initial value is arbitrary. $\nu^{\prime}(0)$ is 0. This choice ensures that there is no singularity in the origin. One gets the value of $p_F^{\prime}(0)$ from $\nu^{\prime}(0)$ by means (\ref{Pf}). Only the central Fermi momentum, $p_F(0)$, remains free parameter.

\begin{figure}
\includegraphics[width=5cm]{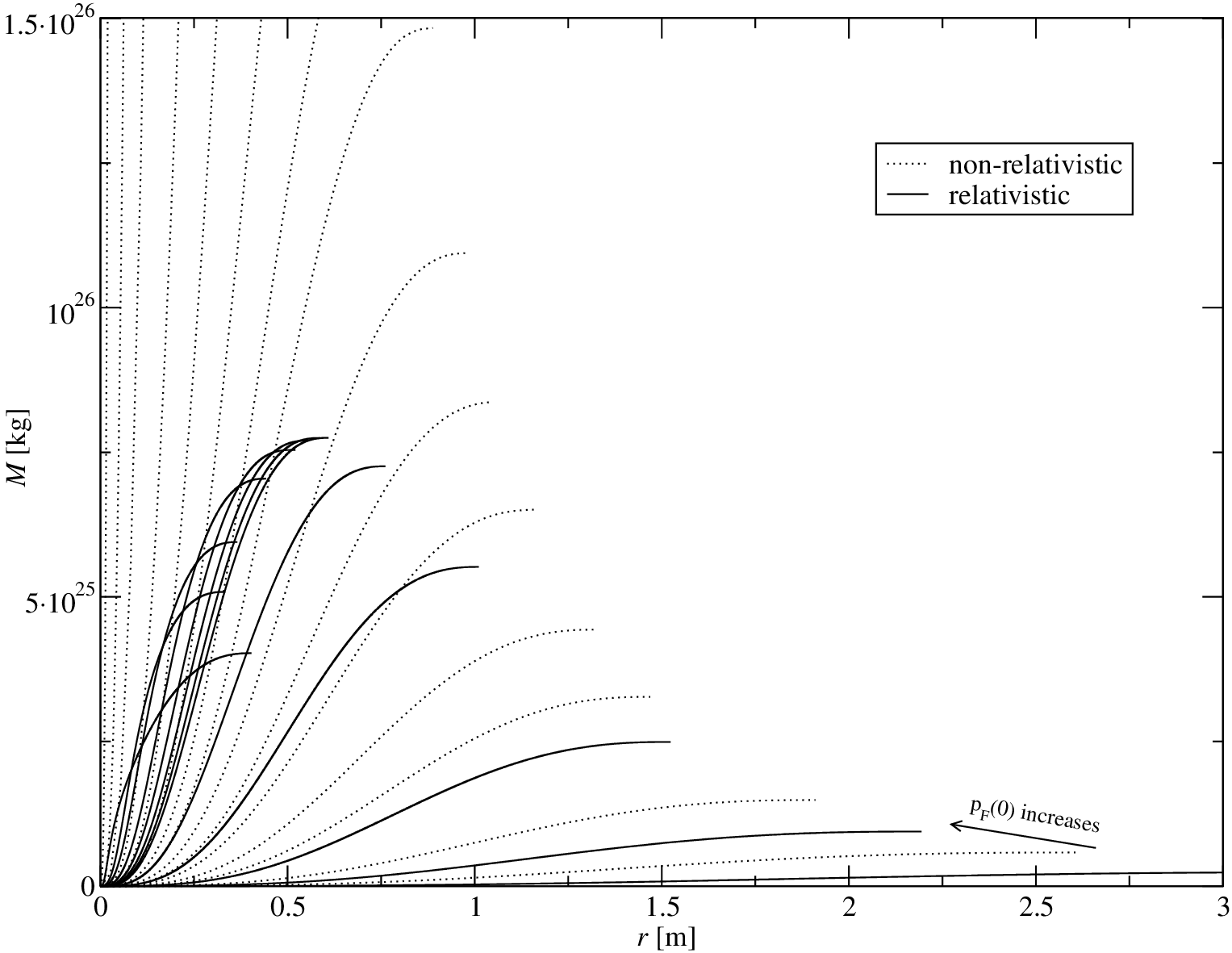}
\caption{\label{fig:Mr}Comparison of relativistic and non-relativistic calculations for mass as function of the distance from the centre for different central Fermi momentum. $m_{0}=10^{11} \rm eV$, Dirac case. }
\end{figure}

We can see the $M(r)$ function of solutions for different $p_F$ in FIG.\ref{fig:Mr}. Solid curves belong to the relativistic case and dotted curves to the non-relativistic one. (We do not depict relativistic and non-relativistic curves with same central $p_F$ for sake of better perspicuity but we have tested that relativistic curves converge to the non-relativistic ones if $p_F(0) \rightarrow 0$.)

As one increases the value of $p_F(0)$, the end point of the non-relativistic curves start from the right bottom corner of the FIG.\ref{fig:Mr} and evolve to the left upper corner. 

\begin{figure}[hb!]
\includegraphics[width=5cm]{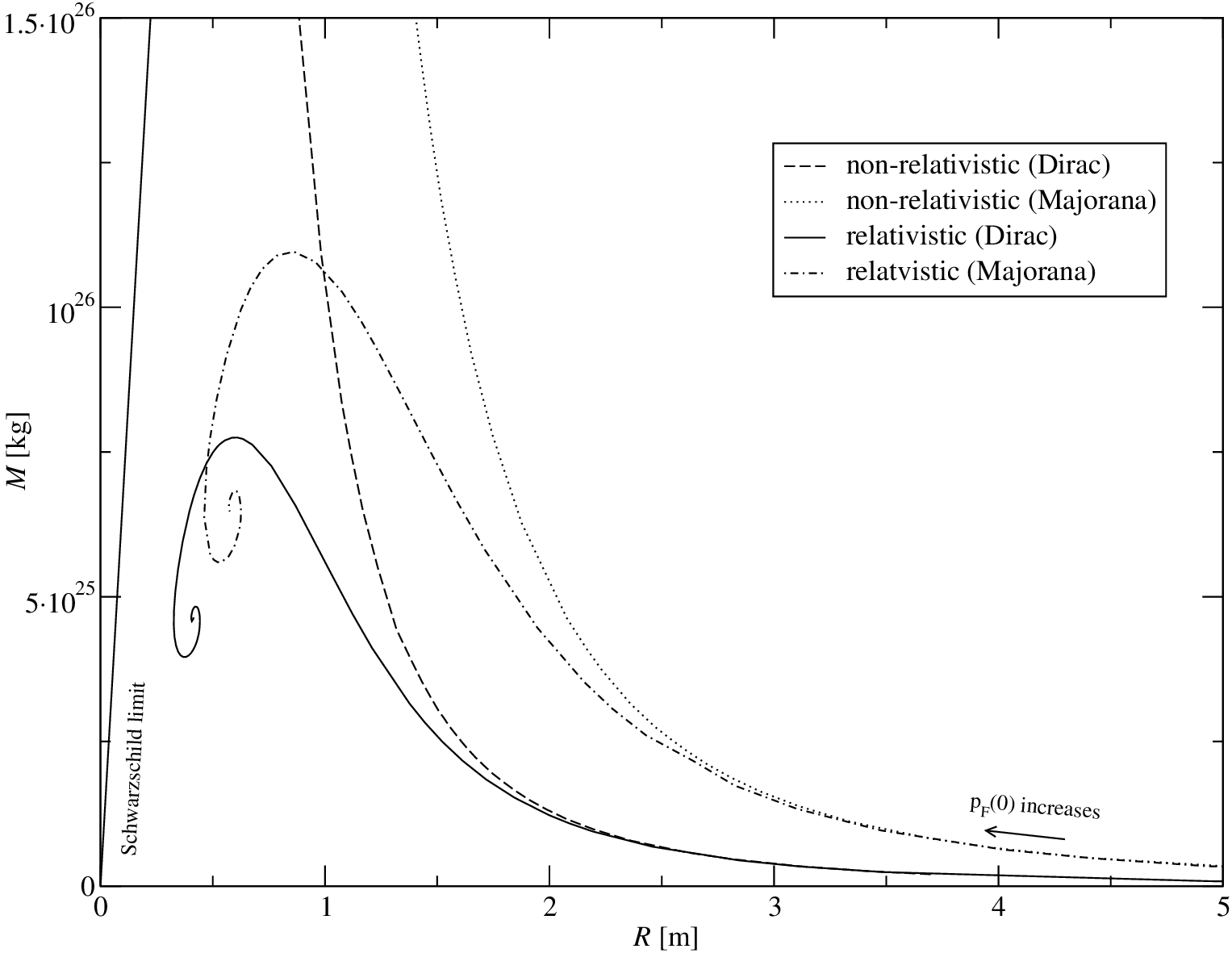}
\caption{\label{fig:MR} Radius of surface of our objects against their total mass.   $m_{0}=10^{11} \rm eV$.}
\end{figure}

FIG.\ref{fig:MR} shows the envelope of the end points of the curves in the FIG.\ref{fig:Mr}. In non-relativistic case one can get solutions for arbitrary high masses but not in relativistic case. We can observe a "turning point" in the relativistic curve. There is a maximum mass here above which there are not solutions of the system. We denote this maximum mass with $M_{max}$ and its location with $R_{max}$.
The ratio $M_{max}/M_{Sch}$ is $\sim 5.29$. 

(One can observe an interesting spiral structure in this plot. We followed this "whirl" till the numerical resolution allowed. This structure means several solutions with same mass. May be infinite for one $M$ value if the spiral makes infinite number of revolution. Of course we can state similar for $R$.)

\begin{figure}
\includegraphics[width=5cm]{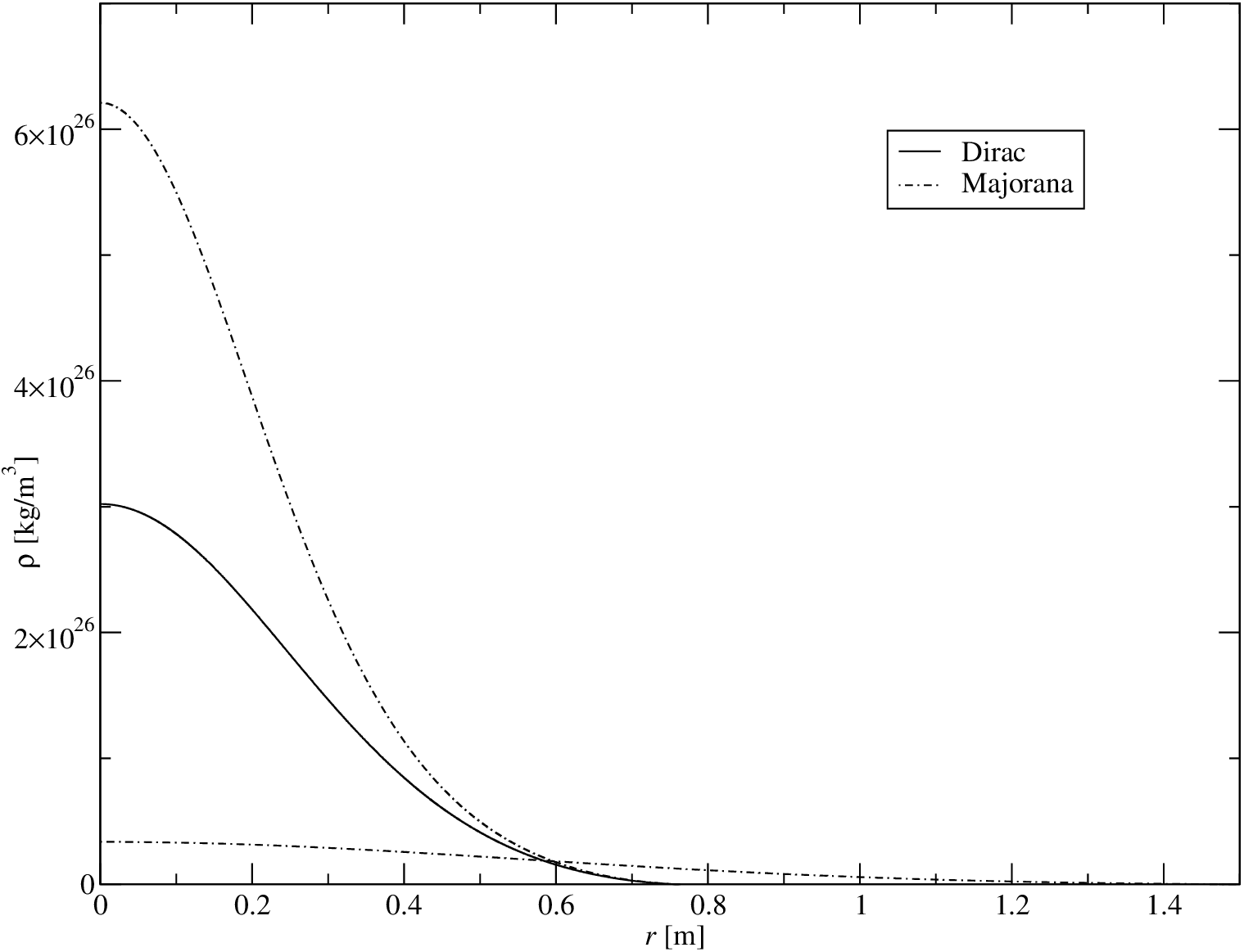}
\caption{\label{fig:DM} Comparison of the density profile of Dirac and Majorana case. $m_{0}=10^{11} \rm eV$.} 
\end{figure}

FIG.\ref{fig:DM} shows the difference between the density distribution of such systems in Dirac and Majorana case. We chose one of the Dirac curves ending in a given point $(R^{Dirac},M^{Dirac})$. There is no Majorana curve which ends in same point but we could depict two corresponding Majorana curves. One of them ends at $R^{Dirac}$ and the other at $M^{Dirac}$. The curve ending at $M^{Dirac}$ spreads to larger volume than its Dirac counterpart because there is no so many energy states in the phasespace as in Dirac case, that is the central density is lower in this case. The curve ending at $R^{Dirac}$ contains more mass than its Dirac counterpart because this higer mass can "press" the Majorana matter to the smaller volume. It is the qualitative reason why the Majorana curve in FIG.\ref{fig:MR} is shifted to the right and up. Quantitatively one can get the Majorana curve in FIG.\ref{fig:MR} from the Dirac one if $r$ and $M$ are multiplied  by a factor $\sqrt{z_{Dirac}/z_{Majorana}} = \sqrt{2}$. 
 
If a fermionic ball reaches the $R_{max}$ with $M>M_{max}$, then there is no static soultion, that is the gravitational collapse is unavoidable. We can conclude that this $M_{max}$ is the mass limit which we have been looking for. If we take into account the scaling feature, we have $M_{max}\approx  0.39 \frac{1}{m_0^2}$ in Dirac case and $M_{max}\approx  0.55 \frac{1}{m_0^2}$ in Majorana case ($M$ is in units of solar mass and $m_0$ in GeV). 

\section{Influence of the annihilation on the masslimit -- Lower bound of the possible maximal mass}

There are several papers about the annihilation cross section of weakly interacting particles which can be candidates of dark matter. One can read a good review about this topic in Bertone's report \cite{Bertone} (see also the references therein). We study the effect of the annihilation on the upper mass limit now.

We can expand the velocity averaged cross section in a Taylor series: $\langle\sigma v \rangle = a + bv^2 + \ldots$ We do not know the value of the factors. One can only estimate upper limits of the total $\langle\sigma v \rangle$ in a given range of $m_0$. We use Beacom's  estimations \cite{Beacom}. We calculate the final mass of such objects after 13.7 Gyear (age of the universe) with these upper limits started from the $M_{max}$ obtained in the previous section. It gives lower bound on the limit mass. (We have to note that it is not a lower bound on the mass of these objects because it is, of course, zero.)

\begin{figure}
\includegraphics[width=5cm]{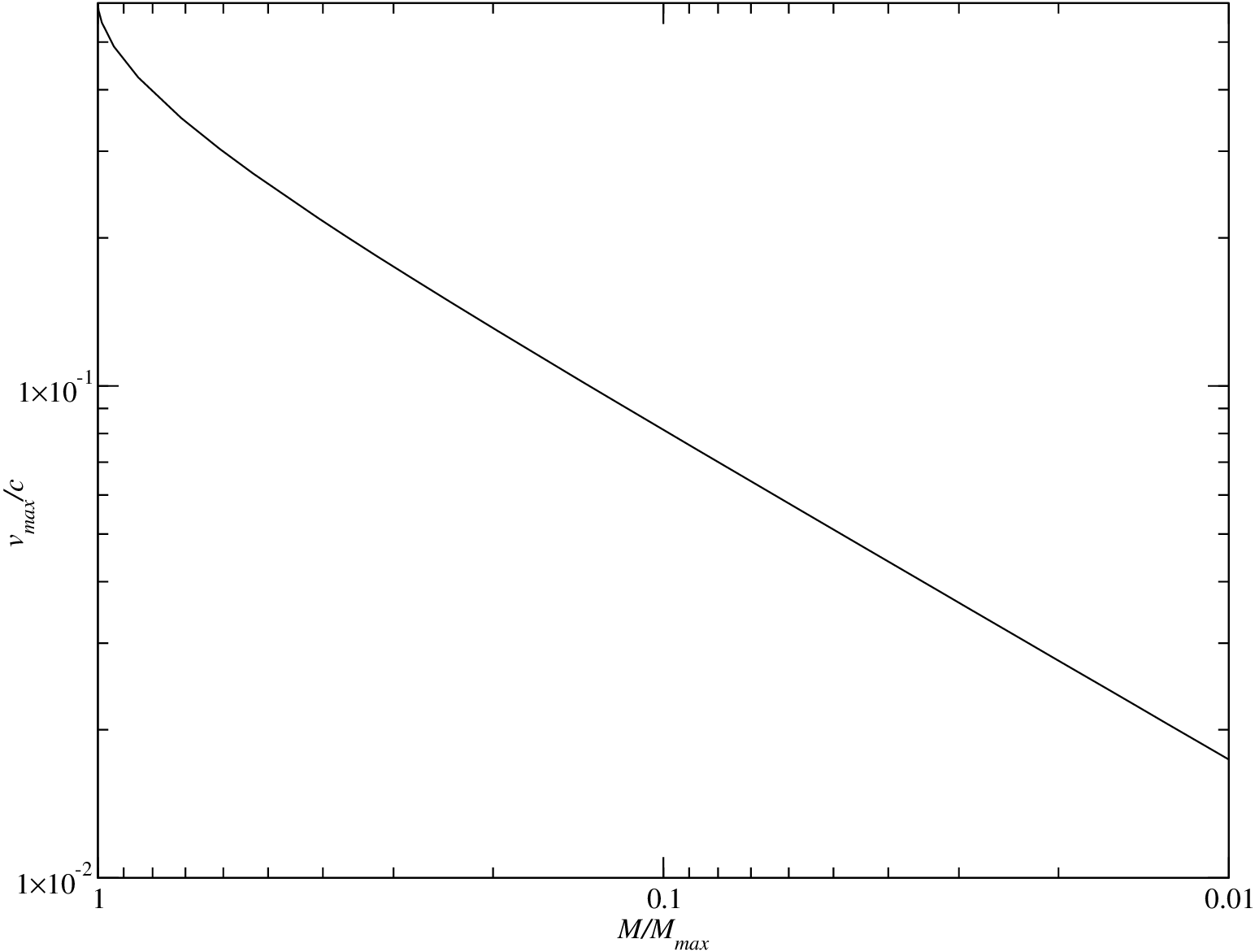}
\caption{\label{fig:pFM} Ratio of the maximal velocity calculated from $p_F$ to speed of light against the ratio of $M$ to $M_{max}$.} 
\end{figure}

We do not make big error in the calculations of the decaying time and the final mass probably if we use the $\langle\sigma v \rangle = constant$ non-relatvistic approximation. 
It is known that $M(t)/M_0=\frac{1}{\sqrt{1+t/\tau}}$ in non-relativistic case \cite{Viollier1}, where $M_0$ is the initial mass and $\tau$ is the characteristic time of the decaying ($\tau \propto M_0^{-2}$). It means that the decaying is very fast at the beginning and slower later. If we used higher terms in the expression of cross section, we just would accelerate the first phase but it would not influence the speed of the latter, non-relatvistic decay phase. 
FIG.\ref{fig:pFM} shows the ratio of the maximal velocity obtained from $p_F$ to speed of light against the ratio of the mass to the maximal mass from relativistic calculation. One can see that the maximal velocities of the particles decrease fast with decreasing $M$. We will see that the final mass is far below the hundredth of the original mass. It justifies our approximation.

One can write the mass loss by the annihilation with these assumptions in the following form:
\begin{equation} \label{Mscale}
\frac{dM}{dt} = - \frac{32\pi \langle\sigma v \rangle}{z} \int_0^{R}\varrho n r^2 dr \equiv - \langle\sigma v \rangle B M^3.
\end{equation}
This equation can be integrated easily:
\begin{equation} \label{Mt}
M(t) = \frac{M_0}{\sqrt{1+2\langle\sigma v \rangle  M_0^{2}Bt}}.
\end{equation}
If we use the scaling feature (\ref{mscaling}), start the mass from $M_{max}$ and use Beacom's upper limits, then we get the following:
\begin{equation} \label{Main}
M(t,m_0) = \frac{M_{max}^{*}\left(\frac{m^{*}_0}{m_0}\right)^{2}}{\sqrt{1+2\langle\sigma v \rangle_{upper} (m_0)  M_{max}^{*2}\left(\frac{m^{*}_0}{m_0}\right)^{4} Bt}},
\end{equation}
where $M_{max}^{*}$ is the maximal mass at a given $m^{*}_0$ and $\langle\sigma v \rangle_{upper}(m_0)$ denotes that $\langle\sigma v \rangle$ depends on $m$. If we set $t=1.37\cdot 10^{10}$ year, then we can get the recent lower limit on the maximal mass of these Fermion balls. 

\begin{figure}
\includegraphics[width=5cm]{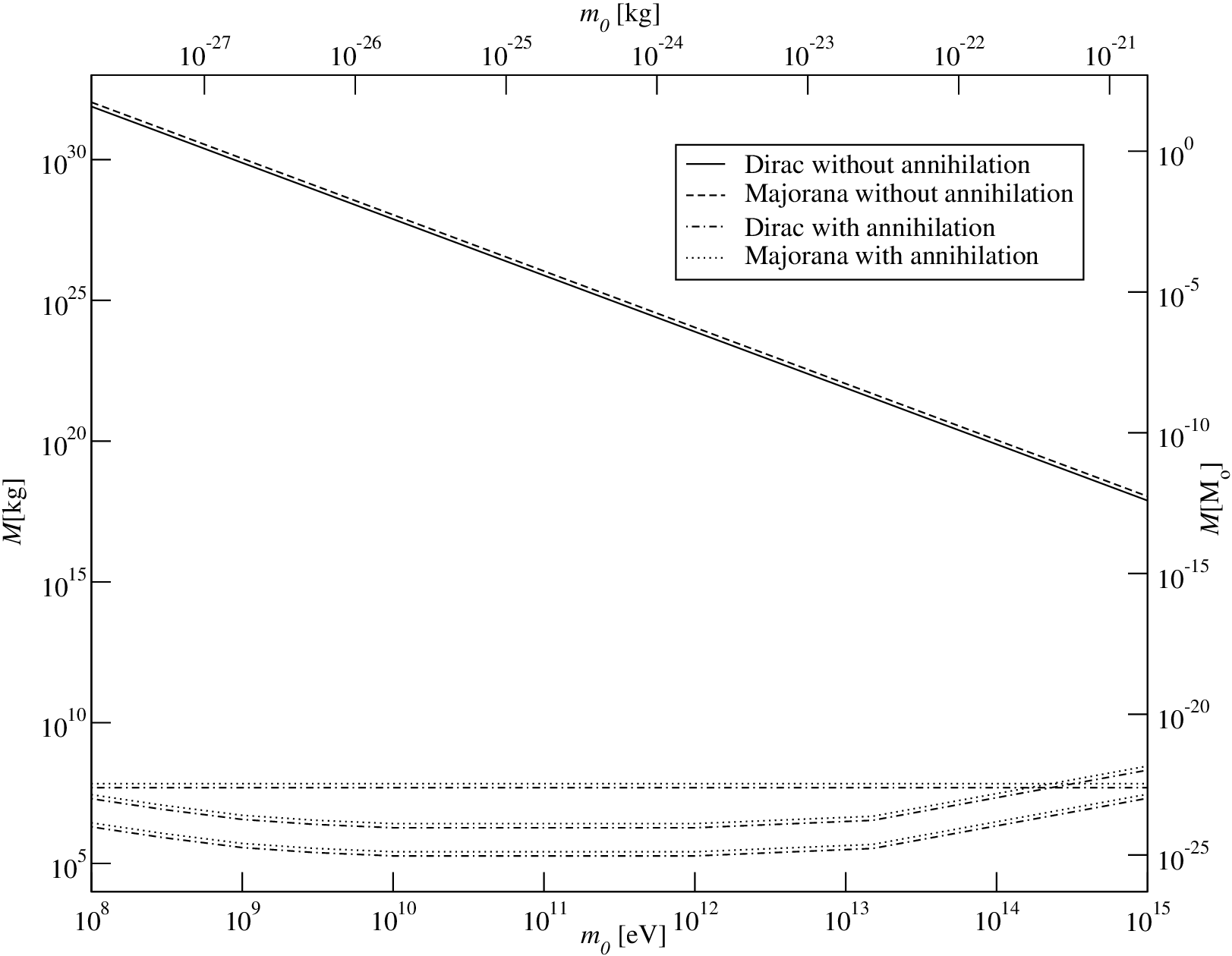}
\caption{\label{fig:mM} Possible maximal mass $M$ of the Fermion balls against the rest mass $m_0$ of Fermions.} 
\end{figure}

We summarize our result in FIG.\ref{fig:mM}. The upper two curves show the upper limit of the limit mass if there is not annihilation. The lowest two curved lines show the mass limit after $1.37\cdot 10^{10}$ year annihilation started from the upper curves, two curved lines above these two belong to $1.37\cdot 10^{8}$ year annihilation. We depict them because Beacom's estimations of $\langle\sigma v \rangle$ give values for the cross section in the late universe. One can observe that even after this shorter time the mass limit is well below the limit without annihilation. We know further that if the  $\langle\sigma v \rangle$ is larger than the natural limit, two horizontal lines, then the dark matter can not be thermal relic.  

Unfortunately the lower limit not strict but we can state this now. Probably the real cross section is much below the limits given by Beacom. If someone can give more strict upper bound on the cross section, our lower limit would increase. 

\section{Conclusions}

We studied self-gravitating systems of weakly interacting fermions. We determined the limit mass above which systems undergo gravitational collapse. We could see that this limit mass is a function of the rest mass of the particles: $M=constant/m^2$. However, this relation was known before we give the exact value of this $constant$ which depends on the type of fermions, namely whether they are Dirac or Majorana spinors. In Majorana case the central density cannot be as high as in Dirac case at same total mass. One can conclude from this that the value of this $constant$ is higher in Majorana case than in Dirac case. It is an upper or absolute masslimit of such objects. 

We calculated the lower limit of the limit mass used the recent upper bound on annihilation cross section of dark matter particles. We could see that this limit is not strict but we can not state more on the base of the recent knowledge.

Further investigations are needed to answer the question what is the probability that such objects gather mass above the mass limit and undergo gravitational collapse or how much dark matter they could  "burn" if they could not collapse.

\end{document}